\documentclass{article}
\usepackage{amssymb}

\begin{document}

\title{Spin correlations and magnetic susceptibilities of 
lightly doped antiferromagnets}
\author{F. Carvalho Dias$^{*}$ and I. R. Pimentel 
  \\ {\small Department of Physics
and CFTC, University of Lisbon} 
   \\ {\small Av. Prof. Gama Pinto 2, 1649-003 LISBOA, Portugal}
   \\ {\small (dated: November 19, 2004) }}
\date{}
\maketitle

\begin{abstract}
We calculate the spin correlation function and the magnetic 
longitudinal and
transverse susceptibilities of a two-dimensional antiferromagnet 
doped with a small concentration of holes, in 
the t-J model. We find that the motion of holes generates spin 
fluctuations which add to the quantum fluctuations, the spin 
correlations decaying with the inverse
 of the spin distance, while increasing with doping as the 
critical hole concentration, where the long-range order 
disappears, is approached. 
Moreover, the longitudinal susceptibility 
becomes finite in the presence of doping, due to the strong 
damping effects induced by the hole motion, while the transverse
 susceptibility is renormalized by softening effects. Both the 
longitudinal and the transverse susceptibilities increase with 
doping, the former more significantly than the latter. Our 
results imply that doping destroys the long-range order while
local antiferromagnetic spin correlations persist.
This is consistent with experiments on the doped copper 
oxide superconductors.


\end{abstract}

\newpage

Since their discovery,$^{1}$ the copper oxide high-$T_{c}$ 
superconductors have shown unusual magnetic characteristics, 
along with the unconventional transport properties.$^{2}$ The 
undoped materials, e.g., La$_{2}$CuO$_{4}$, are antiferromagnetic 
(AF) insulators, and doping, e.g., in 
La$_{2-\delta }$Sr$_{\delta }$CuO$_{4}$, introduces 
holes, which are the charge carriers, in the spin lattice of the 
copper oxide planes. The CuO$_{2}$ planes are described by a 
spin-1/2 Heisenberg antiferromagnet on a square lattice with 
moving holes that strongly interact with the spin array. A 
remarkable feature of the copper oxides is the strong dependence 
of their magnetic properties on the hole concentration $\delta $.
 In previous work$^{3-5}$ we studied the effects of doping on 
various magnetic properties, and showed that the motion of holes 
generates significant softening and damping of the spin 
excitations, leading, in particular, to the disappearance of the 
long-range AF order at a small hole concentration, due to the 
decay of spin-waves. We found that the staggered magnetization 
vanishes at a hole concentration well below the one for which 
the spin-wave velocity vanishes, or even the one for which all 
spin-waves become overdamped. This suggests that although the 
long-range order has disappeared, strong AF correlations persist,
 which allow the spin-wave excitations to exist. This is in 
agreement with experiments in the copper oxides, which show that,
 although the long-range order disappears, AF correlations 
persist up to fairly high doping, into the superconducting 
state.$^{2,6-9}$ It is therefore of interest to study the 
spin correlations in these materials, because of their unusual 
behavior and their possible connection to high-$T_{c}$ 
superconductivity.

In this work we use the $t-J$ model to calculate the spin 
correlation function of a two-dimensional antiferromagnet as a 
function of the hole concentration, which allows to investigate 
the local spin 
fluctuations, and also calculate the longitudinal and transverse 
magnetic susceptibilities, which reflect the global response of 
the system, accounting for the total spin fluctuations. We
 consider zero temperature and the low doping regime where the 
long-range AF order still exists. It is shown that the motion 
of holes generates spin fluctuations which add to the quantum 
fluctuations of the system, and increase with the hole 
concentration. Moreover, we find that the longitudinal spin 
susceptibility, which is zero in a pure Heisenberg 
antiferromagnet at zero temperature, becomes finite in the 
presence of doping, increasing significantly with the hole 
concentration, more pronouncedly than the corresponding 
transverse spin susceptibility.

We describe the copper oxide planes with the t-J model, 
\begin{equation}
H_{t-J}=-t\sum_{<i,j>,\sigma }(c_{i\sigma }^{\dagger }c_{j\sigma
}+h.c.)+J\sum_{<i,j>}\left( \mathbf{S}_{i}\cdot \mathbf{S}_{j}-
\frac{1}{4}n_{i}n_{j}\right) ,  \label{tJ}
\end{equation}
where $\mathbf{S}_{i}=\frac{1}{2}c_{i\alpha }^{\dagger }
\mathbf{\sigma }_{\alpha \beta }c_{i\beta }$ is the electronic 
spin 
operator, $\mathbf{\sigma }$ are the Pauli matrices, 
$n_{i}=n_{i\uparrow }+n_{i\downarrow }$
and $n_{i\sigma }=c_{i\sigma }^{\dagger }c_{i\sigma }$. To 
enforce no 
double occupancy of sites, we use the slave-fermion Schwinger 
boson representation$^{10}$ for the electronic operators 
$c_{i\sigma }=f_{i}^{\dagger }b_{i\sigma }$, where the 
slave-fermion operator $f_{i}^{\dagger }$ creates a hole and the 
boson operator $b_{i\sigma }$ accounts for the spin, subject to 
the local constraint 
$f_{i}^{\dagger }f_{i}+b_{i\uparrow }^{\dagger }
b_{i\uparrow}+b_{i\downarrow }^{\dagger }b_{i\downarrow }=2S$. 
For the undoped system, the model (1) describes a spin-1/2 
Heisenberg antiferromagnet, exhibiting long-range N\'{e}el order 
at zero temperature. The N\'{e}el state is represented by a 
condensate of Bose fields $b_{i\uparrow }=\sqrt{2S}$ and 
$b_{j\downarrow }=\sqrt{2S}$, respectively, in the up and down 
sub-lattices, and the bosons $b_{i}=b_{i\downarrow }$ and 
$b_{j}=b_{j\uparrow }$ are then
spin-excitation operators on the N\'{e}el background. After a
Bogoliubov-Valatin transformation on the boson Fourier transform,
$b_{\mathbf{k}}=u_{\mathbf{k}}\beta _{\mathbf{k}}+v_{\mathbf{k}}
\beta _{-\mathbf{k}}^{\dagger }$, where $u_{\mathbf{k}}=
\left[ \left( (1-\gamma _{\mathbf{k}}^{2})^{-1/2}+1\right) /2
\right]^{1/2} $ and $v_{\mathbf{k}}=-sgn(\gamma _{\mathbf{k}})
\left[ \left(
(1-\gamma _{\mathbf{k}}^{2})^{-1/2}-1\right) /2\right] ^{1/2}$, 
with 
$\gamma_{\mathbf{k}}=(\cos k_{x}+\cos k_{y})/2$, we arrive at the
 effective Hamiltonian 
\begin{equation}
H=-\frac{1}{\sqrt{N}}\sum_{\mathbf{q},\mathbf{k}}f_{\mathbf{q}}
f_{\mathbf{q}-\mathbf{k}}^{\dagger }\left[ V(\mathbf{q},
-\mathbf{k})
\beta _{-\mathbf{k}}+V(\mathbf{q}-\mathbf{k},\mathbf{k})
\beta _{\mathbf{k}}^{\dagger }\right]+\sum_{\mathbf{k}}
\omega _{\mathbf{k}}^{o}\beta _{\mathbf{k}}^{\dagger}
\beta_{\mathbf{k}},
\end{equation}
where, $V(\mathbf{q},\mathbf{k})=zt(\gamma _{\mathbf{q}}
u_{\mathbf{k}}+\gamma_{\mathbf{q}+\mathbf{k}}v_{\mathbf{k}})$ 
represents the interaction between holes and spin-waves resulting
 from the motion of holes with emission and absorption of 
spin-waves, $\omega _{\mathbf{k}}^{o}=(zJ/2)(1-
\gamma _{\mathbf{k}}^{2})^{1/2}$ is the dispersion for spin-waves
 in the undoped antiferromagnet, and $z$ is the lattice 
coordination number ($z=4$), $N$ the number of sites in each 
sub-lattice. The sums are performed in the first Brillouin zone 
of an antiferromagnet on a square lattice.

The magnetic properties are calculated in terms of the spin-wave 
Green's functions 
\[
\begin{array}{l}
D^{-+}(\mathbf{k},t-t^{\prime })=-i<\mathcal{T}
\beta _{\mathbf{k}}(t)
\beta _{\mathbf{k}}^{\dagger }(t^{\prime })> 
\\ 
D^{+-}(\mathbf{k},t-t^{\prime })=-i<\mathcal{T}
\beta _{-\mathbf{k}}^{\dagger}(t)\beta _{-\mathbf{k}}
(t^{\prime })> 
\\ 
D^{--}(\mathbf{k},t-t^{\prime })=-i<\mathcal{T}
\beta _{\mathbf{k}}(t)
\beta_{-\mathbf{k}}(t^{\prime })> 
\\ 
D^{++}(\mathbf{k},t-t^{\prime })=-i<\mathcal{T}
\beta _{-\mathbf{k}}^{\dagger}(t)\beta _{\mathbf{k}}^{\dagger }
(t^{\prime })> ,
\end{array}
\]

\noindent where $<\cdots >$ represents the average over the 
ground state. The spin-wave Green's functions verify the Dyson 
equations 
\[
D^{\mu \upsilon }(
\mathbf{k},\omega )=D_{o}^{\mu \upsilon }(\mathbf{k},\omega )+
\sum_{\alpha\gamma }D_{o}^{\mu \alpha }(\mathbf{k},\omega )
\Pi ^{\alpha \gamma }(\mathbf{k},\omega )D^{\gamma \upsilon }
(\mathbf{k},\omega ), 
\]
with $\mu ,\upsilon
,\alpha ,\gamma =\pm $. The free Green's functions are 
$D_{o}^{-+}(\mathbf{k},\omega )=1/(\omega 
-\omega _{\mathbf{k}}^{o}+
i\eta )$, $D_{o}^{+-}(\mathbf{k},\omega )=1/(-\omega -
\omega _{\mathbf{k}}^{o}+i\eta )$, ($\eta \rightarrow0^{+}$), 
$D_{o}^{--}(\mathbf{k},\omega )=D_{o}^{++}(\mathbf{k},\omega )=0$. 
$\Pi ^{\alpha \gamma }(\mathbf{k},\omega )$ are the self-energies
 generated by the interaction between holes and spin-waves, 
which we calculate in the self-consistent Born approximation 
(SCBA). This corresponds to considering only ''bubble'' diagrams 
with dressed hole propagators, describing the decay of spin-waves
 into ''particle-hole'' pairs. The spin-wave self-energies take 
the form$^{4}$ 
\begin{equation}
\Pi ^{\alpha \gamma }(\mathbf{k},\omega )=\frac{1}{N}
\sum_{\mathbf{q}
}U^{\alpha \gamma }(\mathbf{k},\mathbf{q})\left[ Y(\mathbf{q},
-\mathbf{k} ;\omega )+Y(\mathbf{q}-\mathbf{k},\mathbf{k};-\omega)
\right] , 
\end{equation}

\noindent with $U^{--}(\mathbf{k},\mathbf{q})=U^{++}(\mathbf{k},
\mathbf{q})=V(\mathbf{q},-\mathbf{k})V(\mathbf{q}-\mathbf{k},
\mathbf{k})$, $U^{+-}(\mathbf{k},\mathbf{q})=V(\mathbf{q}
-\mathbf{k},
\mathbf{k})^{2}$, $U^{-+}(\mathbf{k},\mathbf{q})=
V(\mathbf{q},-\mathbf{k})^{2}$ and 
\[
Y(\mathbf{q},-\mathbf{k};\omega )=\int_{0}^{+\infty }
d\omega ^{\prime
}\int_{-\infty }^{0}d\omega ^{\prime \prime }\frac{\rho 
(\mathbf{q},\omega^{\prime })\rho (\mathbf{q}-\mathbf{k},
\omega ^{\prime \prime })}{\omega+\omega ^{\prime \prime }
-\omega ^{\prime }+i\eta } . 
\]
The SCBA provides a spectral function for the holes,$^{9-15}$ 
$\rho (\mathbf{q},\omega )$, which is composed of a coherent 
quasiparticle
 peak with weight $a_{o}\simeq (J/t)^{2/3}$ and dispersion 
$\varepsilon _{\mathbf{q}}\simeq\varepsilon _{\min }+(\mathbf{q}-
\mathbf{q}_{i})^{2}/2m$, with effective mass $m\simeq 1/J$,
 the Fermi surface for the holes consisting of 
pockets, of radius $q_{F}=\sqrt{\pi \delta }$, located at 
$\mathbf{q}_{i}=(\pm \frac{\pi }{2},\pm \frac{\pi }{2})$ in the
Brillouin zone,  and an 
incoherent continuum taking the approximate form 
$h\theta (|\omega |-zJ/2)\theta(2zt+zJ/2-|\omega |)$, with 
$h\simeq (1-a_{o})/2zt$. We calculated the
self-energies to lowest order in the hole concentration 
$\delta $.

The spin correlation function is defined as 
\begin{equation}
C(\mathbf{r})=\frac{1}{2N}\sum_{j}\left( \left\langle 
\mathbf{S}_{j}\cdot 
\mathbf{S}_{j+\mathbf{r}}\right\rangle -\left\langle 
\mathbf{S}_{j}
\right\rangle \cdot \left\langle \mathbf{S}_{j+\mathbf{r}}\right
\rangle \right) , 
\end{equation}

\noindent where the sum runs over all lattice sites. 
Writing the spin operators, 
$S_{j}^{z}$, $S_{j}^{x}=(S_{j}^{+}+S_{j}^{-})/2$, 
$S_{j}^{y}=(S_{j}^{+}-S_{j}^{-})/2i$, in terms of the electron 
operators, one has, $S_{j}^{z}=(c_{j\uparrow }^{\dagger }
c_{j\uparrow }-c_{j\downarrow}^{\dagger }c_{j\downarrow })/2$, 
$S_{j}^{+}=c_{j\uparrow }^{\dagger}c_{j\downarrow }$, 
$S_{j}^{-}=c_{j\downarrow }^{\dagger }c_{j\uparrow }$,
which, using the Schwinger boson representation and the boson 
condensation associated with the N\'{e}el state, leads to 
$S_{j}^{z}=(1-\delta)(1/2-b_{j}^{\dagger }b_{j})$, 
$S_{j}^{+}=(1-\delta )b_{j}$, $S_{j}^{-}=(1-\delta )
b_{j}^{\dagger }$,
 for the up sub-lattice, and $S_{j}^{z}=-(1-\delta )(1/2-b_{j}
^{\dagger }b_{j})$, $S_{j}^{+}=(1-\delta)b_{j}^{\dagger }$, 
$S_{j}^{-}=(1-\delta )b_{j}$, for the down sub-lattice, 
having done the approximation $f_{j}f_{j}^{\dagger }=1-\delta $. 
In terms of the spin-excitation boson operators one has

\begin{eqnarray}
C(\mathbf{r}) &=&(1-\delta )^{2}\frac{1}{4N}\left[ \sum_{j\in 
S(\uparrow)}\left( \left\langle b_{j}b_{j+\mathbf{r}}^{\dagger }
\right\rangle +\left\langle b_{j}^{\dagger }b_{j+\mathbf{r}}
\right\rangle +2\left\langle b_{j}^{\dagger}b_{j}b_{j+\mathbf{r}}
^{\dagger }b_{j+\mathbf{r}}\right\rangle \right. 
\right.  \nonumber 
\\
&&-2\left. \left\langle b_{j}^{\dagger }b_{j}\right\rangle \left
\langle b_{j+\mathbf{r}}^{\dagger }b_{j+\mathbf{r}}\right\rangle 
\right) +\sum_{j\in S(\downarrow)}\left( \left\langle b_{j}
^{\dagger }b_{j+\mathbf{r}}\right\rangle +\left\langle b_{j}
b_{j+\mathbf{r}}^{\dagger }\right\rangle \right.  \nonumber 
\\
&&\left. +2\left. \left\langle b_{j}^{\dagger }b_{j}
b_{j+\mathbf{r}}
^{\dagger }b_{j+\mathbf{r}}\right\rangle -2\left\langle b_{j}
^{\dagger }b_{j}\right\rangle
\left\langle b_{j+\mathbf{r}}^{\dagger }b_{j+\mathbf{r}}\right
\rangle \right) \right] .
\label{corr2}
\end{eqnarray}
After Fourier transform and the Bogoliubov-Valatin 
transformation, 
we make the mean-field decoupling 
$\left\langle ABCD\right\rangle \approx 
\left\langle AB\right\rangle \left\langle CD\right\rangle +
\left\langle AC\right\rangle \left\langle BD\right\rangle +
\left\langle AD\right\rangle \left\langle BC\right\rangle $. 
This allows to express the correlation function (5) in terms of 
the spin-wave Green's functions in the form 
\begin{eqnarray}
C(\mathbf{r}) &=&(-1)^{x+y}(1-\delta )^{2}\left\{ 
\frac{1}{N}\sum_{\mathbf{k}}
(u_{\mathbf{k}}^{2}+v_{\mathbf{k}}^{2})
\cos (\mathbf{k}\cdot \mathbf{r})
\right.
  \label{corr3} \\
&&\times \left[1- 
\int_{0}^{+\infty }\frac{d\omega }{2\pi }
\left( 2\mathrm{Im}D^{+-}(\mathbf{k},\omega )
 +4u_{\mathbf{k}}v_{\mathbf{k}}\mathrm{Im}D^{--}
(\mathbf{k},\omega )\right)\right]
 \nonumber \\
&&\left. +\frac{4}{N^{2}}\sum_{\mathbf{k}_{1},\mathbf{k}_{2}}
\int_{0}^{+\infty }\frac{d\omega _{1}}{2\pi }\int_{0}^{+\infty }
\frac{d\omega _{2}}{2\pi }\cos \left( (\mathbf{k}_{1}-
\mathbf{k}_{2})\cdot \mathbf{r}\right)\right.  
\nonumber \\
&&\left. \times \left[ v_{\mathbf{k}_{1}}^{2}-
(u_{\mathbf{k}_{1}}^{2}+
v_{\mathbf{k}_{1}}^{2})\mathrm{Im}D^{+-}
(\mathbf{k}_{1},\omega _{1})-2u_{\mathbf{k}_{1}}
v_{\mathbf{k}_{1}}\mathrm{Im}D^{--}
(\mathbf{k}_{1},\omega _{1})\right]\right. 
 \nonumber \\
&&\left. \times \left[ u_{\mathbf{k}_{2}}^{2}-
(u_{\mathbf{k}_{2}}^{2}+v_{\mathbf{k}_{2}}^{2})
\mathrm{Im}D^{+-}(\mathbf{k}_{2},\omega _{2})-
2u_{\mathbf{k}_{2}}v_{\mathbf{k}_{2}}\mathrm{Im}D^{--}
(\mathbf{k}_{2},\omega _{2})\right]\right\} ,
\nonumber  
\end{eqnarray}
where $\mathbf{r}=(x,y)$. The pre-factor of $(-1)$ arises when the
 correlation is between sites on different sub-lattices.

To lowest order in the hole concentration, $\delta $, we obtain 
for the correlation function (\ref{corr3}), the expression 
\begin{equation}
C(\mathbf{r})=(1-\delta )^{2}\left[ C_{o}(\mathbf{r})+C_{\delta }
(\mathbf{r})\right]  \label{corr4}
\end{equation}
where 
\begin{eqnarray}
C_{o}(\mathbf{r})&=&\frac{1}{2N}\sum_{\mathbf{k}}
\left( u_{\mathbf{k}}^{2}+v_{\mathbf{k}}^{2}\right) 
\cos (\mathbf{k}\cdot \mathbf{r})
 \\
&& +\frac{1}{N^{2}}\left(\sum_{\mathbf{k}}
u_{\mathbf{k}}^{2}\cos(\mathbf{k}
\cdot \mathbf{r})\right)
\left(\sum_{\mathbf{k}}v_{\mathbf{k}}^{2}
\cos(\mathbf{k}\cdot \mathbf{r})
\right)
\nonumber
\end{eqnarray}
is the correlation function for a pure Heisenberg 
antiferromagnet, accounting for the quantum fluctuations, 
and 
\begin{eqnarray}
C_{\delta }(\mathbf{r}) &=&-\left[ 1+\frac{1}{N}\sum_{\mathbf{k}}
\left( u_{\mathbf{k}}^{2}+
v_{\mathbf{k}}^{2}\right)
\cos (\mathbf{k}\cdot \mathbf{r}) \right]
\frac{1}{N}\sum_{\mathbf{k}}
\cos (\mathbf{k}\cdot \mathbf{r})   
 \\
&& \times
\left[ u_{\mathbf{k}}v_{\mathbf{k}}\frac{\mathrm{Re}
\Pi ^{++}(\mathbf{k},\omega _{\mathbf{k}}^{o})}{\omega 
_{\mathbf{k}}^{o}}-2u_{\mathbf{k}}v_{\mathbf{k}}
\int_{0}^{+\infty }\frac{d\omega }{\pi}\frac{\mathrm{Im}\Pi ^{++}
(\mathbf{k},\omega )}{\omega ^{2}-(\omega _{\mathbf{k}}^{o})^
{2}}\right.  
\nonumber \\
&&\left. +(u_{\mathbf{k}}^{2}+v_{\mathbf{k}}^{2})\int_{0}^
{+\infty }\frac{d\omega }{\pi }\frac{\mathrm{Im}\Pi ^{-+}
(\mathbf{k},\omega )}{(\omega +\omega _{\mathbf{k}}^{o})^{2}}
\right]
\nonumber
\end{eqnarray}
contains the effect of doping on the spin correlations 
associated to the hole motion; the pre-factor $(1-\delta )^{2}$ 
corresponds to spin dilution, being negligible in the low doping 
regime considered. In Fig.1 we present the correlation function 
$C(r)$, Eq.(\ref{corr4}), calculated for two different
directions, $x=y$ and $y=0$, in the case of $\delta=0.02$, and 
the pure case, $\delta=0.0$. One sees that the spin correlations are 
independent of the spatial direction, a result that is verified 
at any doping. In Fig.2 we plot the correlation function 
as a function of the spin distance $r$, 
for various hole concentrations. We 
observe that $C(r)$ increases with doping, and decays, at large 
distances, as $1/r$ (Fig.2 inset), both in the pure and in the doped
 cases. One can describe the behavior of $C(r)$ at large 
distances as $C(r)=A(\delta)/r$,
where $A(\delta)=A_{o}+B\delta^{\alpha}$, with 
$A_{o}=1 / \sqrt{2} \pi$ and $\alpha=0.42$. 
$A(\delta)$ contains the doping dependence, which is represented 
in Fig.3. The dominant 
contribution to $C_{\delta}(r)$, at large $r$, 
comes from the imaginary part of the spin-wave self-energies, 
which depend on the hole concentration essentially as 
$\sqrt{\delta}$.$^{5}$
In Fig.4 we compare the increase of $C(r)$ with the 
hole concentration $\delta$ at fixed small $r$ (Fig.4a) and large 
$r$ (Fig.4b). The decay of $C(r)$ with $1/r$ was expected 
for the undoped case since in a two dimensional antiferromagnet
 at zero temperature the correlation length is infinite.$^{2,17}$
 One also expected $C(r)$ to increase with doping since the 
motion of holes generates spin fluctuations which eventually 
lead to the destruction of the long-range AF order at
 a finite critical concentration $\delta_{c}$. In previous 
work,$^{5}$ we found that the staggered magnetization vanishes at
 a small critical concentration (e.g. $\delta_{c}\simeq 0.07$ 
for $t/J=3$), 
while the long-wavelength spin excitations remain well defined up
 to a higher hole concentration ($\delta^{*}\simeq 0.17$ also 
for $t/J=3$). Here, we find 
that the doping does not qualitatively change the behavior of 
$C(r)$ with $r$, as compared to the pure case, which reflects the 
robustness of the local AF order in the doped material. Spin 
correlations in the copper oxides were 
studied before, both experimentally$^{2,6-9}$ and theoretically,
$^{18-20}$ but in a higher doping regime where the long-range 
AF order has already disappeared. In this
regime the spin correlations decrease with increasing doping, 
as the system moves away from the critical hole concentration.

In the presence of long-range AF order one distinguishes a 
longitudinal and a transverse susceptibility. The longitudinal 
spin susceptibility is defined as 
\begin{equation}
\chi _{\Vert }=\chi _{\Vert }(\mathbf{k}=0,\omega =0) ,
\end{equation}
where the dynamical susceptibility is given by 
\[
\chi _{\Vert }(\mathbf{k},\omega )=i\int_{0}^{+\infty }dte^{i\omega t}
\left
\langle\left[ S^{z}(\mathbf{k},t),S^{z}(-\mathbf{k},0)\right] 
\right\rangle .
\]
In terms of the spin-wave Green's functions one has 
\begin{eqnarray*}
\chi _{\Vert } &=&\lim_{\mathbf{k}\rightarrow 0}i\frac{1}{N}
\sum_{\mathbf{k}_{1}}\int_{-\infty }^
{+\infty }\frac{d\omega }{2\pi }
\left[ 2u_{\mathbf{k}_{1}}v_{\mathbf{k}_{1}}u_{\mathbf{k}_{1}-
\mathbf{k}}v_{\mathbf{k}_{1}-\mathbf{k}}-u_{\mathbf{k}_{1}}^{2}
u_{\mathbf{k}_{1}-\mathbf{k}}^{2}-v_{\mathbf{k}_{1}}^{2}
v_{\mathbf{k}_{1}-\mathbf{k}}^{2}\right] 
\\
&&\times D^{+-}(\mathbf{k}_{1},\omega )D^{-+}(\mathbf{k}_{1}
-\mathbf{k},-\omega )
\end{eqnarray*}
which to lowest order in the hole concentration gives 
\begin{equation}
\chi _{\Vert }=4\frac{1}{N}\sum_{\mathbf{k}}
\int_{-\infty }^{+\infty }\frac{d\omega }{2\pi }
\frac{\mathrm{Im}\Pi ^{+-}
(\mathbf{k},\omega )}{(\omega-\omega _{\mathbf{k}}^{o})^{3}} .
\end{equation}

The transverse spin susceptibility is defined by

\begin{equation} 
\chi _{\bot }=\chi _{\bot }(\mathbf{k}=0,\omega =0) , 
\end{equation}
where 
\[
\chi _{\bot }(\mathbf{k},\omega )=i\int_{0}^{+\infty }dte^{i\omega t}\left
\langle
\left[ S^{x}(\mathbf{k},t),S^{x}(-\mathbf{k},0)\right] \right
\rangle .
\]
In terms of the spin-wave Green's functions the transverse spin
susceptibility is expressed as$^{3}$
\[
\chi _{\bot }=-\lim_{\mathbf{k}\rightarrow 0}\left( \frac{1-
\gamma _{\mathbf{k}}}{1+\gamma _{\mathbf{k}}}\right) ^{1/2}
\left[ \mathrm{Re}D^{+-}(\mathbf{k},0)+\mathrm{Re}D^{++}
(\mathbf{k},0)\right] ,
\]
which, to lowest order in the hole concentration 
$\delta $, is given by 
\begin{equation}
\chi _{\bot } =\lim_{\mathbf{k}\rightarrow 0}\frac{1}{zJ(1+
\gamma _{\mathbf{k}})}\left[ 1-\frac{2}{zJ(1-
\gamma _{\mathbf{k}}^{2})
^{1/2}}\left[ \mathrm{Re}\Pi ^{+-}(\mathbf{k},0)
+\mathrm{Re}\Pi ^{++}(\mathbf{k},0)\right]\right] .
\end{equation}
We found that $\chi_{\bot}$ takes the simple form
\[
\chi_{\bot}=Z_{\chi}\chi_{\bot}^{o}
\]
where $\chi _{\bot }^{0}=1/(2zJ)$ is the 
transverse spin susceptibility for a pure Heisenberg 
antiferromagnet, and 
$Z_{\chi}=1+4\delta a_{o}^{2}\left( \frac{t}{J}\right)^{2}$
is a renormalization factor.

Comparing Eqs. $(11)$ and $(13)$ one sees that the motion of 
holes influences the longitudinal and transverse susceptibilities
 in different ways, the former is produced by the imaginary part
 of the self-energy, while the latter is renormalized by the real
 part of the self-energies. In a pure Heisenberg antiferromagnet 
the longitudinal susceptibility is zero. However, with doping 
$\chi _{\Vert }$ 
acquires a finite value due to the decay of spin-waves into 
''particle-hole'' pairs, generated by hole motion. The 
renormalization of $\chi_{\bot}$ reflects a softening of the 
spin coupling induced by the hole motion. In Figs. 5 and 6 we 
plot the longitudinal, Eq.(11), and the transverse,
Eq.(13), susceptibilities as a function of the hole 
concentration for $t/J=3,4$, in the doping range where 
the long-range AF order
 exists, in the approach considered.$^{5}$ We find that both 
susceptibilities increase with doping, though the longitudinal 
one is far more sensitive to the hole concentration than the 
transverse one. The transverse susceptibility reflects the 
stiffness of the antiferromagnetic lattice. In contrast, the 
longitudinal susceptibility is set by the strong damping effects, 
which are also responsible for the 
disappearance of the long-range AF order at low 
doping.$^{5}$ When the long-range order is broken, the 
susceptibility of the system should be essentially given by
$\chi = \frac{1}{3}\chi_{\Vert}+\frac{2}{3}\chi_{\bot}$, with the
 longitudinal susceptibility providing an important contribution.
 Also, in the ceramic samples whose crystal axis are randomized 
the susceptibility $\chi$ is given by an average of the 
susceptibilities for the three directions. An increase of the 
spin susceptibility with doping has in fact been observed 
experimentally.$^{21-24}$

In summary, we studied the effects of hole motion on the spin 
correlation function and the magnetic longitudinal and 
transverse susceptibilities of a two-dimensional antiferromagnet 
doped with a small concentration of holes. We found that the spin
 fluctuations increase with doping, the spin correlations decaying
 with the inverse of the spin distance, which indicates that the 
local AF correlations remain quite robust. 
Furthermore, we show that the longitudinal magnetic 
susceptibility acquires a finite value in the presence of doping,
 due to the strong damping effects generated by the hole motion, 
while the transverse magnetic susceptibility is renormalized. 
Both susceptibilities show a significant increase with doping, 
which is however more pronounced in the longitudinal one.
Our results imply that doping destroys the long-range AF order
while local spin correlations persist. 
This is consistent with experimental observations in 
the copper oxide high-$T_{c}$ superconductors.

\bigskip

\noindent $^{*}$Electronic address: fdias@cii.fc.ul.pt

\bigskip

\noindent {\LARGE References:}

\noindent $^{1}$J. G. Bednorz and  K. A. M\"{u}ller, Z. Phys. B 
\textbf{64}, 189 (1986).

\noindent $^{2}$M. A. Kastner, R. J. Birgeneau, G. Shirane, and 
Y. Endoh, Rev. Mod. Phys. \textbf{70}, 897 (1998).

\noindent $^{3}$I. R. Pimentel and R. Orbach, Phys. Rev. B 
\textbf{46}, 2920 (1992).

\noindent $^{4}$I. R. Pimentel, F. Carvalho Dias, L. M. Martelo, 
and R. Orbach, Phys. Rev. B \textbf{60}, 12329 (1999).

\noindent $^{5}$F. Carvalho Dias, I. R. Pimentel, and R. Orbach, 
Phys. Rev. B \textbf{61}, 1371 (2000).

\noindent $^{6}$R. J. Birgeneau et al, 
Phys. Rev. B \textbf{38}, 6614 (1988).

\noindent $^{7}$B. Keimer, R.J. Birgeneau, A. Cassanho, Y. Endoh, 
R. W. Erwin, M. A. Kastner, and G. Shirane, Phys. Rev. Lett. \textbf{67}, 
1930 (1991).

\noindent $^{8}$B. Keimer, N. Belk, R. J. Birgeneau, A. Cassanho, 
C. Y. Chen, M. Greven, M. A. Kastner, A. Aharony, Y. Endoh,
R. W. Erwin, and G. Shirane, Phys. Rev. B \textbf{46}, 
14034 (1992).

\noindent $^{9}$G. Shirane, R. J. Birgeneau, Y. Endoh, and
M. A. Kastner, Physica B \textbf{197}, 158 (1994).

\noindent $^{10}$G. Martinez and P. Horsch, Phys. Rev. B 
\textbf{44}, 317 (1991).

\noindent $^{11}$C. L. Kane, P. A. Lee, and N. Read, 
Phys. Rev. B \textbf{39}, 6880 (1989).

\noindent $^{12}$F. Marsiglio, A. E. Ruckenstein, S. Schmitt-Rink,
 and C. M. Varma, Phys. Rev. B \textbf{43}, 10882 (1991).

\noindent $^{13}$Z. Liu and E. Manousakis, Phys. Rev. B 
\textbf{45}, 2425 (1992).

\noindent $^{14}$E. Dagotto, Rev. Mod. Phys. \textbf{66}, 
763 (1994).

\noindent $^{15}$N. M. Plakida, V. S. Oudovenko, and V. Yu. 
Yushankhai, Phys. Rev. B \textbf{50}, 6431 (1994).

\noindent $^{16}$B. Kyung and S. I. Mukhin, Phys. Rev. B 
\textbf{55}, 3886 (1997).

\noindent $^{17}$R. Manousakis, Rev. Mod. Phys. \textbf{63}, 
1 (1991).

\noindent $^{18}$O. P. Sushkov, Phys. Rev. B \textbf{54}, 
9988 (1996).

\noindent $^{19}$S. Winterfeldt and D. Ihle, Phys. Rev. B 
\textbf{58}, 9402 (1998).

\noindent $^{20}$A. Sherman and M. Schreiber, Eur. Phys. J. B 
\textbf{32}, 203 (2003).

\noindent $^{21}$J. B. Torrance, A. Bezinge, A. I. Nazzal, 
T. C. Huang, S. S. P. Parkin, D. T. Keane, S. J. LaPlaca, 
P. M. Horn, and G. A. Held, Phys. Rev. B \textbf{40}, 
8872 (1989).

\noindent $^{22}$Y.-Q. Song, Mark A. Kennard, 
K. R. Poeppelmeier, and W. P. Halperin, Phys. Rev. Lett. 
\textbf{70}, 3131 (1993).

\noindent $^{23}$T. Nakano, M. Oda, C. Manabe, N. Momono,
Y. Miura, and M. Ido, Phys. Rev. B \textbf{49}, 16000 (1994).

\noindent $^{24}$S. Ohsugi, Y. Kitaoka, and K. Asayama, 
Physica C \textbf{282-287}, 1373 (1997).

\newpage

\noindent {\large Figure Captions:}

\medskip

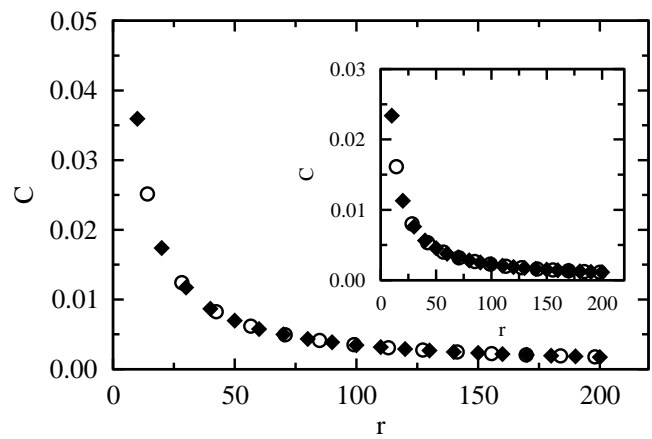
\noindent FIG. 1. Correlation function $C(r)$  \textsl{vs} spin 
distance $r$, at the hole concentration $\delta=0.02$, for the 
directions $x=y$ (open circles) and $y=0$ (diamonds) on the 
square lattice, with $t/J=3$. Inset: $C(r)$ \textsl{vs} $r$, in 
the pure antiferromagnet ($\delta=0.0$).

\smallskip

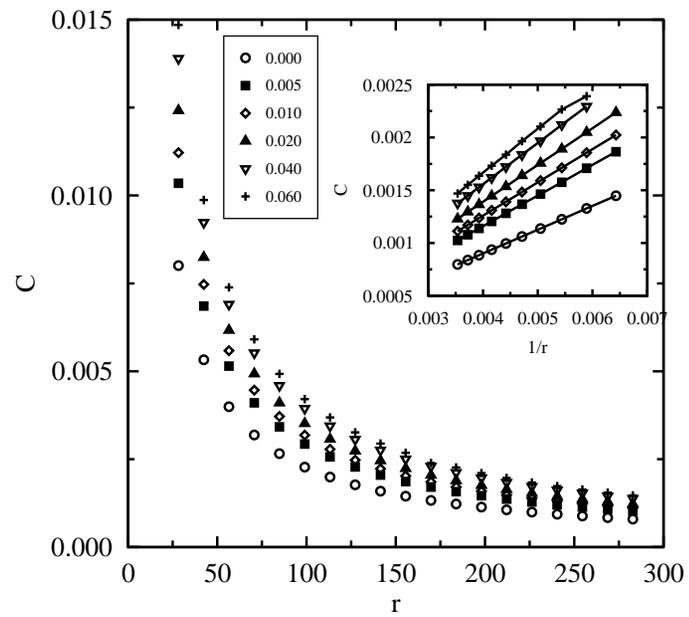
\noindent FIG. 2. Correlation function $C(r)$ \textsl{vs} 
spin distance $r$, $x=y$, for various hole concentrations $\delta$, 
with $t/J=3$. Inset: $C(r)$ \textsl{vs} $1/r$ for, large $r$.

\smallskip

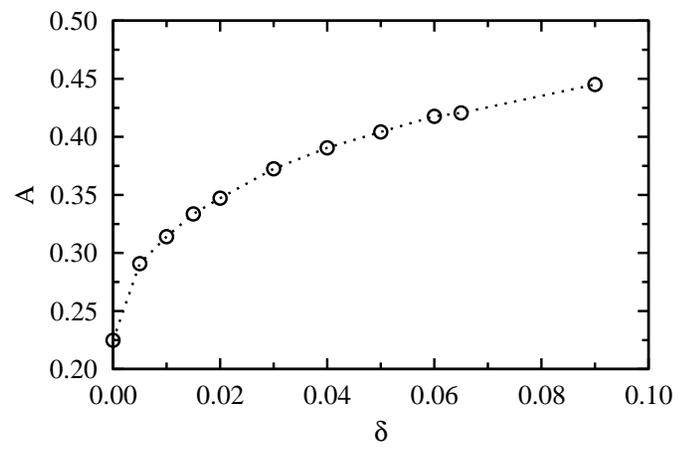
\noindent FIG. 3. Spin correlation amplitude $A(\delta)$, 
at large spin distances, \textsl{vs} hole concentration 
$\delta$, with $t/J=3$. 

\smallskip

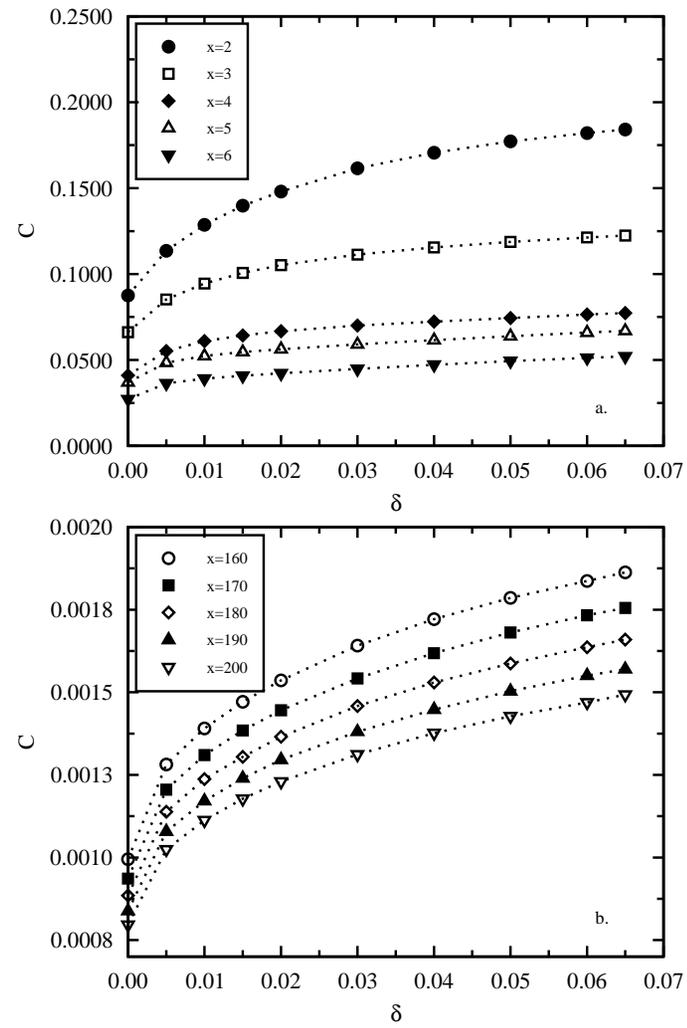
\noindent FIG. 4. Correlation function $C(r)$ \textsl{vs} hole 
concentration $\delta$, with $t/J=3$, for fixed $r$, $x=y$. 
a: in the range of small $r$. b: in the range of large $r$.

\smallskip

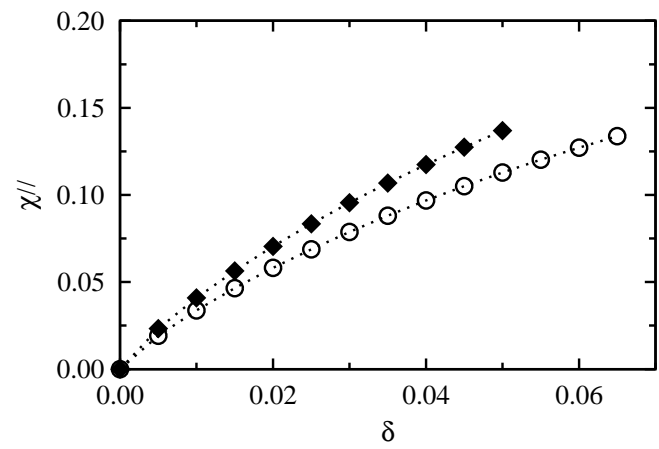
\noindent FIG. 5. Longitudinal susceptibility $\chi_{\Vert}$ as 
a function of doping $\delta$, for $t/J=3$ (open circles) and 
$t/J=4$ (diamonds).

\smallskip

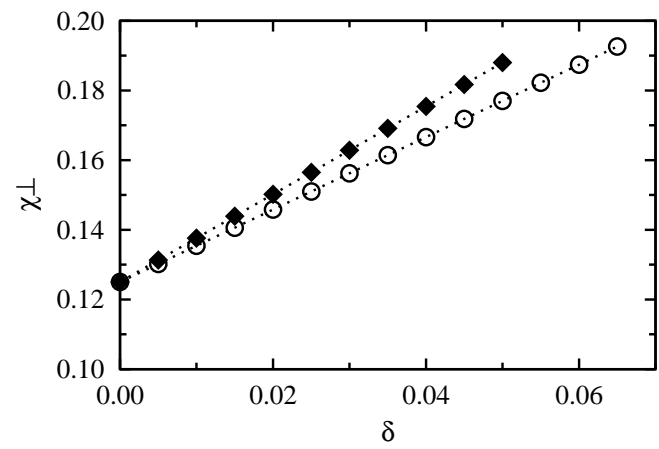
\noindent FIG. 6. Transverse susceptibility $\chi_{\bot}$ as a 
function of doping $\delta$, for $t/J=3$ (open circles) and 
$t/J=4$ (diamonds).

\end{document}